# Single DNG Interface Makes a Better Perfect Lens

Gilad Rosenblatt, Guy Bartal, Meir Orenstein

Department of Electrical Engineering, Technion, Haifa 32000, Israel

meiro@ee.technion.ac.il

**Abstract**. We show that a single-interface between regular and double-negative (DNG) media constitutes the core structure of a perfect lens and, furthermore, substantially outperforms the highly discussed DNG slab-based lens under imperfect conditions, maintaining unbounded resolution. We identify the perfect-lensing mechanism as the excitation of a Brewster (real improper) mode at the interface – a mode with a completely flat-band dispersion at media-matching conditions, perfectly transferring all incident waves including evanescent. A lossless DNG slab retains the single-interface perfect-lensing characteristics solely because such a Brewster mode excited at the front (input) slab interface merges in a non-interacting fashion with a confined surface-mode at the rear (output) – thereby preserving its original flat-band dispersion. While the single interface perfect lens retains its high resolution under loss, imperfect media-matching and off-resonance excitations, the DNG slab lens performance rapidly deteriorates under such conditions due to excitation of the symmetric and anti-symmetric slab surface-modes.

## 1. Introduction

In 1967 Veselago contemplated electromagnetic wave propagation in fictional media having both negative permittivity and permeability. His predictions for such double-negative (DNG) media included left handedness, a negative index of refraction, and reverse Doppler and Cherenkov effects [1]. It nevertheless took three decades for such media to enter the realm of possibility, following a series of works that culminated in the first fabrication [2], and subsequent negative-index measurement [3], of artificial DNG media in the microwave range [4-5].

Meanwhile, a theoretical work by Pendry, predicting that a media-matched (defined as $\mu_2/\mu_1=\varepsilon_2/\varepsilon_1$ where $\varepsilon$ and $\mu$ are the permittivities and permeabilities of two adjacent media 1,2) slab of DNG medium could achieve perfect-lensing by amplifying evanescent waves [6], made significant headways in making such media desirable for various applications – paving the way to the exploding field of metamaterials [7], in topics such as cloaking, super-resolution, transformation-optics, and numerous others [8-11].

This paradigm shifting proposal incited a long-standing debate which has yet to be unequivocally settled by experiment. Some pointed out inconsistencies with Veselago's analysis, yet seconded its conclusions for stationary fields [12]; others disputed the possibility of evanescent wave amplification beyond some finite slab width due to finite energy considerations [13], also claiming the slightest absorption negates any amplification due to alleged singularity in the field solution – to which it was argued absorption only limits resolution non-abruptly (i.e. resolution improves indefinitely with decreased absorption) [14].

More recently, evanescent amplification began to be associated with surface-polariton resonance excitations of the DNG slab [15-18]. This thread could be traced back to Merlin, studying a lossless media-mismatched DNG slab, showing the resultant field fits an either even or odd surface-polariton excitation, and resolution depends logarithmically on mismatch magnitude [16]. Colin arrived at similar conclusions studying near-field excitation in a dispersive DNG slab by incident finite spectral-width pulses, arguing further that at the media-matching frequency, where Pendry's solution is obtained, even and odd surface-polaritons are co-excited, inhibiting coherent reconstruction, though rapidly decaying in lossy media [17]. Conversely, in a recent letter modeling a perfect-lens using two coupled harmonic oscillators, Wee and Pendry argued that a simultaneous excitation of the even and odd surface-polaritons – which cancel at one interface and combine at the other – gives rise to the evanescent amplification inside the slab congenial with perfect-lensing [18]. A similar approach was also taken by Zhang and Khurgin in studying super-resolution in a metal slab [19]. Evidently, the aforementioned association is still being debated: it remains unclear whether surface-polariton excitation is essential for, assists in, or is detrimental to, perfect-lensing in a DNG slab.

Herein we show that the primary perfect-lens operation resides in a single-interface between double-positive (DPS) and DNG materials rather than a DNG slab (two interfaces). The single-interface real-improper mode underlying perfect transmission has a flat dispersion for the media-matched case, and zero reflection (a Brewster mode). Also, the perfect-lensing of a DNG slab is facilitated by the Brewster mode of the slab, and not by the even nor the odd surface polariton modes, nor their superposition and field cancellation.

This comprehension is instrumental in analyzing 'perfect-lensing' efficacy under less stringent conditions, such as loss or media-mismatch, accomplished by studying variations to the corresponding Brewster mode's dispersion and field profile. We show that under such conditions a single-interface perfect-lens vastly outperforms the slab lens, surprisingly maintaining perfect resolution.

## 2. The single-interface flat dispersion solution

The DNG-DPS single-interface supports guided surface-polariton modes in both polarizations [20], which can be either forward or backward propagating depending on permittivity and permeability values [21]. Still, in between those two regimes there exists another possibility: the formation of a completely flat-dispersion eigenmode in both polarizations provided the materials are media-matched.

We derive this flat-band mode using modal analysis with explicit media dispersion. Here, Drude models were selected for DNG permittivity ($\varepsilon_{DNG}$) and permeability ($\mu_{DNG}$) frequency dependencies, as simplified Lorenzians well above resonance (full-fledged Lorenzian models yield similar results):

$$\varepsilon_{DNG} = 1 - \omega_{p,\varepsilon}^2 \big/ \omega\left(\omega + j\omega_{\tau,\varepsilon}\right)$$
$$\mu_{DNG} = 1 - \omega_{p,\mu}^2 \big/ \omega\left(\omega + j\omega_{\tau,\mu}\right) \tag{1}$$

with plasma and scattering angular frequencies for the permittivity and permeability, $\omega_{p,\varepsilon},\omega_{\tau,\varepsilon}$ and $\omega_{p,\mu},\omega_{\tau,\mu}$, respectively ($\varepsilon_{DPS},\mu_{DPS}=1$). A TM-polarized wave solution for the y-directed magnetic field phasor $\mathbf{H}=f(x) \cdot e^{-j\beta z} \cdot \mathbf{y}$, z-propagating along the interface (at $x=0$) with a propagation constant $\beta$ ($e^{j\omega t}$ convention), has an amplitude profile:

$$f(x) = \begin{cases} e^{\kappa_{DNG} x} & , x < 0 \\ e^{-\kappa_{DPS} x} & , x > 0 \end{cases} \tag{2}$$

where $\kappa_{DNG,DPS}=\pm(\beta^2-(\omega/c)^2 \cdot \varepsilon_{DNG,DPS} \cdot \mu_{DNG,DPS})^{1/2}$, and $c$ is the vacuum speed of light. Maxwell boundary constraints dictate its dispersion relation [21]

$$\kappa_{DPS}/\varepsilon_{DPS} + \kappa_{DNG}/\varepsilon_{DNG} = 0 \tag{3}$$

which yields

$$\beta = k_0 \cdot \sqrt{\frac{\varepsilon_r}{\varepsilon_r + 1}} \cdot \sqrt{\frac{\varepsilon_r - \mu_r}{\varepsilon_r - 1}} \qquad (4)$$

where $\varepsilon_r = \varepsilon_{DNG}/\varepsilon_{DPS}$, $\mu_r = \mu_{DNG}/\mu_{DPS}$, and $k_0 = \omega/c \cdot (\varepsilon_{DPS}\mu_{DPS})^{1/2}$. For TE $\varepsilon$ and $\mu$ are interchanged. Notably, two mode types separately conform with this dispersion relation: a regular confined mode where the mode profile (equation (2)) exponentially decays away from the interface ($\kappa_{DNG,DPS} > 0$), and an improper mode where the mode profile exponentially diverges on both sides ($\kappa_{DNG,DPS} < 0$). We dwell upon the importance of this improper mode after first considering the formation of flat dispersion and the resulting perfect-lensing.

The first root in the RHS of equation (4) constitutes a surface-polariton-like component diverging at a resonance frequency for which $\varepsilon_r = -1$, while the second root incorporates both permittivity and permeability dependencies and can be zero if the media are matched ($\varepsilon_r = \mu_r$). This latter term encapsulates the difference between the single-negative (e,g, metal) and double-negative cases.

Figure 1 (lossless case, $\omega_{\tau,\varepsilon} = \omega_{\tau,\mu} = 0$) shows, for several $r_p = \omega_{p,\varepsilon}/\omega_{p,\mu}$ values, the dispersion of forward-propagating waves for $\varepsilon_r < \mu_r$ ($\omega_{p,\varepsilon} > \omega_{p,\mu}$) residing below the resonance frequency, and backward-waves for $\varepsilon_r > \mu_r$ ($\omega_{p,\varepsilon} < \omega_{p,\mu}$) residing above resonance (opposite for TE). In between stands out the media-matched case where the dispersion curve in both polarizations coincide as $\varepsilon_r$ and $\mu_r$ become identical ($\omega_{p,\varepsilon} = \omega_{p,\mu}$; with loss, also $\omega_{\tau,\varepsilon} = \omega_{\tau,\mu}$). In this scenario $\beta$ collapses to zero at all frequencies – aside from a resonant frequency at which the pole counteracts this zero. Strictly speaking, evaluating $\beta(\omega)$ at resonance involves two concurrent limits: nulling the difference between the functions $\varepsilon_r(\omega)$ and $\mu_r(\omega)$ (media-matching), and approaching the resonance frequency. The limiting process (though not the result) depends on the specific frequency dependencies of media parameters, here we use equation (1).

Taking the limit $r_p = \omega_{p,\mu}/\omega_{p,\varepsilon} \to 1$ and deriving the lossless (TM or TE) dispersion curve makes the solution self-evident (figure 1): a completely flat-dispersion curve is formed, whereby at a single resonant frequency every value of propagation constant $\beta$ is supported by the mode in both polarizations, whereas at all other frequencies $\beta = 0$.

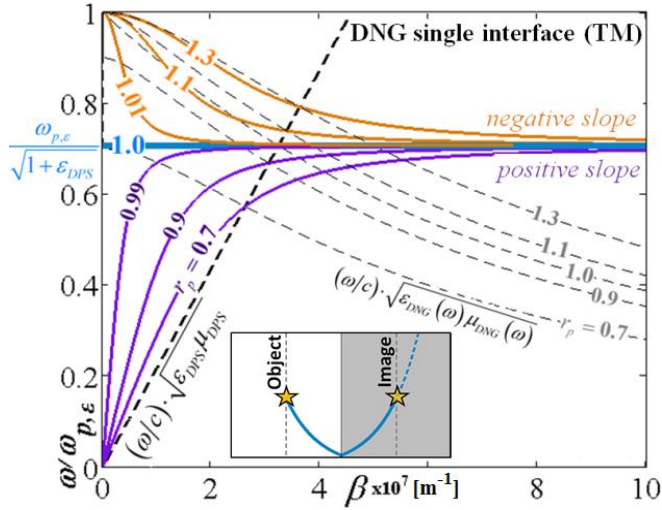

**Figure 1.** TM dispersion curve of a lossless DNG-DPS single-interface (confined as well as improper) mode, for various $r_p$ values – smaller (purple), larger (orange), and equal (blue) to 1 – the latter exhibiting flat dispersion; DPS light-line – dashed black line; DNG light-line for the various $r_p$ values – dashed grey lines. $\omega_{p,\varepsilon} = 1.37 \times 10^{16}$ rad/sec. For TE $r_p \leftrightarrow 1/r_p$; inset: illustration of single-interface perfect-lensing by the improper mode at the flat-band condition.

More rigorously, the multivariate limit evaluation depends on the path of approach such that any $\beta$ value is obtained at the resonance frequency $\varepsilon_r(\omega_{res}) = -1$ (not only $\beta \to \infty$, as in surface-polaritons). For instance, approaching $r_p \to 1, \omega \to \omega_{res}$ (media-matching at resonance) along $\omega = \omega_{res} \cdot r_p^{\alpha}$ with arbitrary $0 \leq \alpha < 1$ yields $\beta \cdot c = \omega_{res}/(1-\alpha)^{1/2}$, namely any $\beta$ value is applicable. This also applies to Lorenzian media dispersion (quadratic dependence) or any sum thereof – i.e. any physically reasonable substitution of lossless material constants generates the flat-band solution.

## 3. Origin and mechanism of DNG perfect lensing and the single-interface perfect lens

The pivotal argument in this study is that (*i*) through the excitation of this single-interface flat-band solution perfect-lensing is materialized in the single-interface, and (*ii*) that the contributing mode is the improper rather than the confined (figure 1 inset). Exciting these incoming flat-band solutions (real improper modes) by a source (object) reveals all merits of perfect-lensing: undistorted transmission of all spatial frequencies with proper 'amplification' of the evanescent fields.

### 3.1. The improper single-interface mode as the source for perfect lensing in a single-interface

It may seem odd, at first glance, that we consider the improper (double-diverging) rather than the regular confined (double-converging) mode. Yet, it is a critical point neglected in perfect-lens analyses: this improper mode is a Brewster mode, related to a zero of the single-interface reflection coefficient *r*, while the confined mode is a reflection pole. Hence, the improper mode describes an incoming wave from a source to the interface, transmitted into the other side without reflection – the exact scenario of the perfect-lens. The confined mode, on the other hand, describes waves emanated from the interface outwards – representing the wrong choice.

The field profile *f(x)* describing reflection (from the DPS side) by a single DPS-DNG interface (*x*=0) has the form (inverted phase sign necessary in the DNG medium)

$$f(x) = \begin{cases} e^{-\kappa_{DPS} x} + r e^{+\kappa_{DPS} x} & x < 0 \\ t e^{+\kappa_{DNG} x} & x > 0 \end{cases} \quad (5)$$

from which the single-interface transmission (*t*) and reflection (*r*) coefficients are extracted

$$t = (2\kappa_{DNG}/\varepsilon_{DNG})/(\kappa_{DNG}/\varepsilon_{DNG} - \kappa_{DPS}/\varepsilon_{DPS})$$
$$r = (\kappa_{DNG}/\varepsilon_{DNG} + \kappa_{DPS}/\varepsilon_{DPS})/(\kappa_{DNG}/\varepsilon_{DNG} - \kappa_{DPS}/\varepsilon_{DPS}) \quad (6)$$

A reflectionless transmission scenario (*r*=0, *t*=1) corresponds to an excitation of the improper single-interface Brewster mode, as equation (5) then fits its double-diverging mode profile and the numerator of *r* equals its dispersion relation. A perfect-lensing scenario (*r(β)*≡0, *t(β)*≡1), when all spatial components undergo reflectionless transmission, therefore corresponds to this Brewster mode being supported for any *β*, i.e. having a flat dispersion.

Furthermore, had we solved here an excitation problem by a line source located in the DPS side (*x*=–*h*), then the field solution (the spatial Green function) between the source and the interface would have the form [22]

$$H_y(x,z) \propto \int_{-\infty}^{\infty} \frac{e^{-\kappa_{DPS}(x+h)} + r(\beta)e^{+\kappa_{DPS}(x-h)}}{\kappa_{DPS}} e^{-j\beta z} d\beta \quad (7)$$

At the perfect lensing condition (*r(β)*≡0) all contributions of regular modes (potential poles of *r*) in equation (7) are eliminated, with only the (improper) Brewster mode contributions left. This clearly indicates that when *r(β)*≡0 the object would not excite regular surface polariton modes. The complete and exact solution for the Green function of this configuration reaffirms these conclusions [22].

The question of divergence at infinity is irrelevant – improper modes are not power-orthogonal and thus couple to radiation eventually, not describing the far-field well. But the perfect-lens is a near-field device and improper modes genuinely describe the near-field in the vicinity of a source, as is well-known in leaky-wave devices [23], and related phenomena in quantum physics [24]. Thus, the exponentially increasing field profile of the improper mode extending beyond the image plane in the DNG material would not contribute to the field solution in the excitation problem, but it will dominate in between the source and image.

*3.2. The perfect lensing mode in a lossless DNG slab*

One may wonder how perfect-lensing is retained in the more complex, yet very familiar, DNG slab configuration. It is not due to the even or odd surface-polariton modes (nor their combinations), which are the confined solutions of the slab. Their dispersion relations (TM, slab thickness *d*) [25-26]

$$\kappa_{DPS}/\varepsilon_{DPS} + \tanh(\kappa_{DNG}d/2)\kappa_{DNG}/\varepsilon_{DNG} = 0 \qquad (8a)$$

$$\kappa_{DPS}/\varepsilon_{DPS} + \coth(\kappa_{DNG}d/2)\kappa_{DNG}/\varepsilon_{DNG} = 0 \qquad (8b)$$

do not generate a flat-band slab mode at the media-matching condition. These slab modes are associated with in-phase (even, equation (8a)) and anti-phase (odd, equation (8b)) coupling of two confined single-interface modes (having flat dispersion), and this coupling is of the interacting type, following an avoided crossing path with splitting of the dispersion curve into those of the even and odd slab modes – which substantially deviate from flat (figure 2).

On equal footing, perfect-lensing cannot be attributed to a combination of such confined slab modes (as suggested in [15,18]) – since each of these modes is excitable at a single frequency and their modal field combinations are far from accommodating with the actual evanescent field transmission through a perfect-lens.

The viable modal solution to perfect-lensing in a slab must correspond to reflectionless uniform transmission. First, it must be a Brewster mode, characterized by a zero rather than a pole in the slab reflection coefficient *R* (from the DPS side). Its field profile is thus of the form (interfaces at $x=\pm l/2$)

$$f_{ns}(x) = \begin{cases} e^{-\kappa_{DPS}(x+d/2)} + Re^{\kappa_{DPS}(x+d/2)} & x < -d/2 \\ F_1 e^{\kappa_{DNG}(x+d/2)} + F_2 e^{-\kappa_{DNG}(x-d/2)} & -d/2 < x < d/2 \\ Te^{\kappa_{DNG}d} e^{-\kappa_{DPS}(x-d/2)} & x > d/2 \end{cases} \qquad (9)$$

with *R*=0 – which satisfies Maxwell's boundary conditions provided $F_2=0, T=F_1=1$; i.e. it has a non-symmetric modal field profile – diverging from one side of the slab while decaying from the other – generated by the combination of the improper (at the front slab interface) and the confined (rear interface) single-interface modes (figure 2, bottom).

Second, it must exhibit flat dispersion. Indeed, this non-symmetric mode satisfies the exact same dispersion relation as a single-interface mode, given by equation (3). The fields of both confined and improper single-interface modes perfectly overlap inside the slab: each mode is unperturbed by the additional interface provided that on the latter the other mode type has evolved. Hence, this combination is non-interacting and maintains the single-interface dispersion curve, exhibiting flat dispersion for matched-media (figure 2).

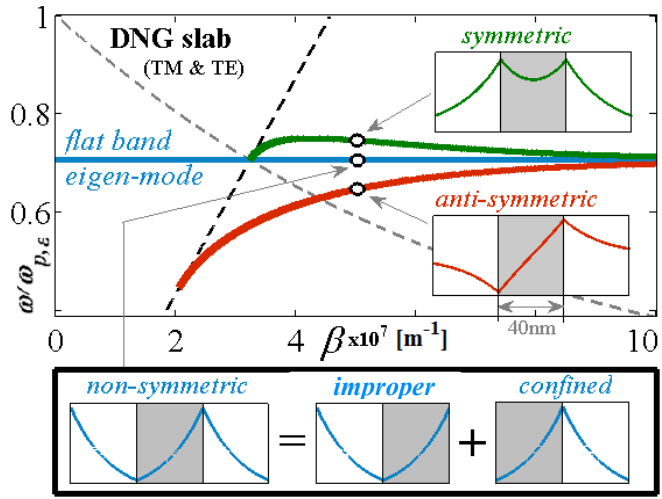

**Figure 2.** Dispersion curves of a lossless media-matched DNG slab (TM and TE coincide): even (green), odd (red) and flat-band (blue) modes. Inset: field profiles at marked points; below: improper and confined single-interface mode-combination constituting the non-symmetric mode. Above (below) resonance frequencies, involve even (odd) mode excitation. $\omega_{p,\varepsilon}=\omega_{p,\mu}=1.37\times10^{16}$rad/sec.

The association with perfect-lensing is immanent: once excited this mode profile translates a spatial component a $2d$ distance across the slab; its polarization-independent flat dispersion insures all spatial components are unilaterally supported. All spatial frequencies emitted by a source on one side would therefore perfectly match this mode's field profile (for appropriate $\beta$ values) thereby exciting it, enabling their undistorted translation to the image plane.

This association allows us to address several known dilemmas, such as the energy associated with evanescent amplification across the slab. In a perfect-lensing context, such amplitude amplification merely echoes the stationary non-symmetric mode profile, not any energy transfer across the slab; rather, energy should principally propagate along the slab in the eigenmode propagation direction – fundamentally differing from the interpretation of [18], examining an even and odd mode combination in which a beating effect directs energy across.

Specifically, in a flat-band (i.e. zero group velocity), energy propagation along the slab is also halted so this complete scenario is localized and stationary. Actual energy transfer across the slab by evanescent fields must relate to either mode excitation transients (not steady-state) or perturbations exerted by photonic detection schemes at the image plain.

## 4. The robustness of the single-interface perfect lens

In a lossy scenario the resonance condition $\varepsilon_r=\mu_r=-1$ is not fully satisfied as $\varepsilon_r$ and $\mu_r$ have imaginary components – i.e. flat dispersion, and therefore perfect-lensing, seems unattainable even with media-matching (in agreement with [27]). Although the perfect-lensing mechanism in both the single-interface and the slab lenses is related, their performance under loss or frequency offsets widely differs. As we show next, while slab transmission properties heavily deteriorate, the single-interface lens still maintains unbounded and almost uniform transmission – and therein lies its basic and applicative significance.

*4.1. Lensing performance deterioration in a lossy DNG slab and the role of even and odd slab modes*

Excluding gain that exactly compensates loss (regaining the perfect-lensing condition) media-matching in the lossy case nullifies the dispersion curve ($\beta \equiv 0$) of the non-symmetric Brewster mode (given by equation (4)). Nevertheless, this does not entail a complete collapse in resolution: the slab's field amplitude reflection ($R$) and transmission ($T$) coefficients, derived from equation (9) (single-interface reflection coefficient $r$ given by equation (6))

$$R = r\left(e^{2\kappa_{DNG}d}-1\right)\big/\left(1-r^2 e^{2\kappa_{DNG}d}\right)$$
$$T = (1+r)^2 \big/ \left(1-r^2 e^{2\kappa_{DNG}d}\right) \tag{10}$$

reveal otherwise (figure 3). The distinct flat-band formation at near-resonant frequencies (black line in figure 3a) is composed of near-zero-reflection continuum-solutions which approximately have a non-symmetric modal field profile (figure 3d-e). They substitute for the Brewster mode across a wide spatial frequency ($\beta$) band (figure 3d-e), enabling almost reflectionless transmission, which is limited by the loss-related backbending of the even and odd surface-modes (whose near-pole signature is seen in white in figure 3a,b). This backbending represents the $\beta$ value from which the modal field profiles heavily distort (figure 3f) and the transmission plummets – revealing the detrimental role these surface-modes play for lensing. Nevertheless, the 'satellite' continuum modes of the Brewster mode enable high-fidelity super-resolution which steadily improves with diminishing loss (in agreement with [14]).

Likewise, as frequency deviates from resonance these satellite modes enable high fidelity up to increasingly lower spatial components at which the even (above resonance) or odd (below resonance) surface-modes are directly excited – which becomes unavoidable even without loss (fitting well with [16]). In any case, media-matching always marks the optimum, as mismatch leads to deterioration in the flat dispersion of the non-symmetric eigenmode (figure 1), shrinking resolution as the even and odd surface-modes further protrude into the pass-band (in agreement with [17]).

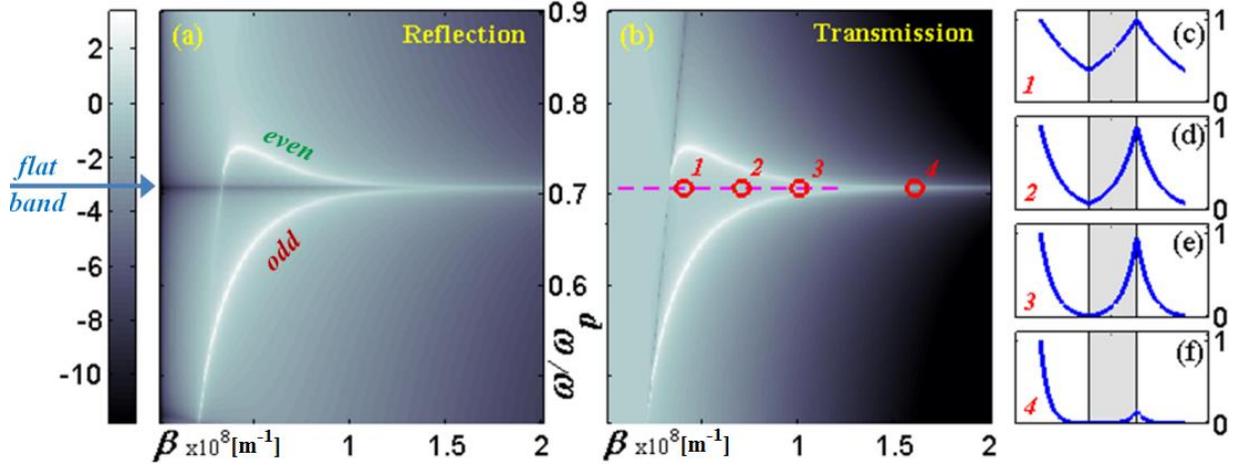

**Figure 3.** $\omega$-$\beta$ dependence of (a) transmission and (b) reflection coefficients (logarithmic colorbar) for a lossy media-matched 30nm thick DNG slab. (c-f) Modal field profiles of four continuum-solutions marked 1-4 in (b). Magenta dashed line in (b): high (>80%) transmission at the almost-resonant frequency – fitting the near-zero-reflection band in (a) reminiscent of the flat-band eigenmode, yet spatially limited by the even and odd mode backbending, where transmission plummets (double-pole) and the field profile distorts. $\omega_{p,\varepsilon}=\omega_{p,\mu}=1.37\times10^{16}$rad/sec, $\omega_{\tau,\varepsilon}=\omega_{\tau,\mu}=4\times10^{13}$rad/sec (material loss at resonance $0.267\mu m^{-1}$).

### 4.2. The superiority of the single-interface perfect lens under realistic conditions

This is where the single-interface perfect-lens substantially outperforms the slab perfect-lens – under imperfect conditions, due to the absence of the 'parasitic' even and odd surface-modes. To support this claim we inspect transmission (from the DPS side) by a single DPS-DNG interface (equation (6)). Figure 4a depicts the transmission of a single-interface lens (material loss at resonance $0.267\mu m^{-1}$, as in figure 3): it retains unbounded flat dispersion (infinite spatial resolution) compared to the slab's modest $\lambda/5$ resolution limit under the same conditions.

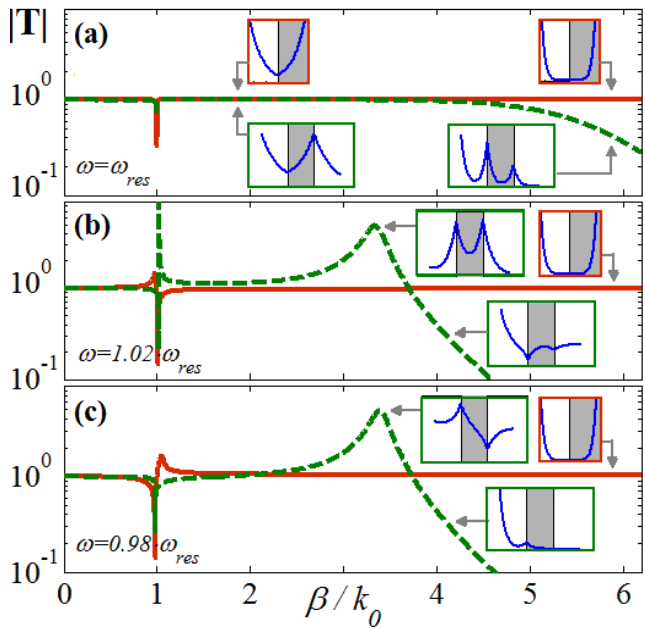

**Figure 4.** Amplitude transmission (logarithmic) of media-matched lossy DNG-DPS single-interface (solid red) and DNG slab (dashed green, $d$=30nm) vs. $\beta$ (normalized by $k_0=\omega_{res}/c$) at three near-resonance frequencies: (a) $\omega_{res}$ (fits the dashed magenta line in Fig. 3b), (b) $0.98\cdot\omega_{res}$, (c) $1.02\cdot\omega_{res}$. Inset: field profile at selected points (arrows) shows deterioration of slab transmission modes due to interference with the even and odd surface-modes, absent for the single-interface, rendering it superior. $\omega_{p,\varepsilon}=\omega_{p,\mu}=1.37\times10^{16}$rad/sec, $\omega_{\tau,\varepsilon}=\omega_{\tau,\mu}=4\times10^{13}$rad/sec.

Another crucial merit of the single-interface lens over the slab is its superior frequency margin (deviation from resonance) for which it maintains undistorted transmission (figure 4b,c): the single-interface lens retains unbounded flat dispersion even for frequency deviation of ±2% (with aforementioned loss) while the slab lens resolution further deteriorates to $\lambda/3$.

The single-interface lens robustness and superiority over the slab can be explained by examining transmission at the asymptotic of large spatial components ($\beta \gg k_0$) to obtain its spatial cut-offs. For the slab, equation (10) then becomes ($\kappa_{DNG,DPS} \approx \beta$)

$$T = \left(\frac{e^{\kappa_{DNG}d}-1}{1-r^{-1}}+1\right)^{-1}\left(\frac{e^{\kappa_{DNG}d}-1}{1+r^{-1}}+1\right)^{-1} \approx \left(\frac{1+\varepsilon_r^{-1}}{2}e^{\beta d}+1\right)^{-1}\left(\frac{1+\varepsilon_r}{2}e^{\beta d}+1\right)^{-1} \qquad (11)$$

Using Bode plot formalism on $|T(e^{\beta d})|$ it is obvious why the slab is a low-pass filter – its transmission has two complex poles at $2/(1+\varepsilon_r^{\pm 1})$ leading to a –40dB/dec attenuation onward from its spatial cut-off

$$\beta_{\text{cut-off}}(\omega) \approx d^{-1}\cdot\ln\left(2/|1+\varepsilon_r(\omega)|\right) \qquad (12)$$

It best performs when $\text{Re}\{\varepsilon_r\}=-1$, i.e. at $\omega=(\omega^2_{l\text{-}res}-\omega^2_{\tau,\varepsilon})^{1/2}$ (where $\omega_{l\text{-}res}=\omega_{p,\varepsilon}/(1+\varepsilon_{DPS})^{1/2}$ is the lossless resonance frequency), when its poles around $\pm j\omega_{l\text{-}res}/\omega_{\tau,\varepsilon}$ lead to a cut-off (in all but the lossless case) at $\beta_{\text{cut-off}}\approx\ln(\omega_{l\text{-}res}/\omega_{\tau,\varepsilon})/d$. For the parameters of figure 4, it yields $\beta_{\text{cut-off}}\approx 5.67\cdot k_0$, while at a ±2% frequency deviation equation (12) yields $\beta_{\text{cut-off}}\approx 3.29\cdot k_0, 3.35\cdot k_0$ – all in agreement with the exact calculations of figure 4. This verifies that the transmission poles, introduced by the even and odd surface-modes (whose dispersion relation constitutes the denominator) severely limit both spatial resolution and frequency bandwidth.

For the single-interface, on the other hand, the situation is quite different. Equation (6) becomes

$$t = 1+r \approx 2/(1-\varepsilon_r) \qquad (13)$$

representing an all-pass filter with a near-1 uniform response over all spatial components – without cut-off due to the absence of poles (i.e. 'parasitic' modes), as is corroborated by figure 4. This implies unlimited resolution even with loss over a wide operational frequency bandwidth – distinguishing the single-interface perfect-lens from all other 'perfect-lenses', and opening a path to realistic application.

## 5. Conclusion

In conclusion, we have shown that perfect-lensing is a characteristic of a *single-interface* between media-matched DPS/DNG. The relegation of perfect-lensing to the more complex slab structure resides solely in the fact that the single-interface modes involved are non-interacting. Lensing performance for the slab is severely deteriorated by loss, media-mismatch, and frequency shifts from resonance, due to the 'parasitic' surface-polariton modes. Conversely, the single-interface lens performs essentially unhindered under such imperfect circumstances – exhibiting unbounded flat-dispersion transmission. It therefore vastly outperforms the slab lens. Its high tolerance to excitation frequency is possibly indicative to its faster perfect-lensing evolution. These characteristics of a single-interface perfect-lens are paramount to actual realization of near-field lensing.